\documentstyle[preprint,aps]{revtex}

\newcommand{\lsim}{\raisebox{0.3mm}{\em $\, <$} 
\hspace{-3.3mm} \raisebox{-1.8mm}{\em $\sim \,$}}

\begin{document}

\draft

\preprint{}

\title{Three-Flavor Analysis of Neutrino Mixing with and without 
Mass Hierarchy\footnote[2]{Talk presented at Inauguration Conference of 
Asia Pacific Center for Theoretical Physics, June 4-10, 1996, Seoul, Korea, 
and at Institute for Nuclear Theory Program, ``{\it Nucleosynthesis in 
Big Bang, Stars and Supernovae}'' (INT-96-2) June 24-August 30, 1996, 
Seattle, WA.}} 

\author{Hisakazu Minakata}
\address{Department of Physics, Tokyo Metropolitan University \\
Minami-Osawa, Hachioji, Tokyo 192-03, Japan}

\date{November, 1996}

\preprint{
\parbox{5cm}{
TMUP-HEL-9613\\
hep-ph/9612259\\
}}

\maketitle

\begin{abstract}
I summarize the results of barely model-dependent phenomenological 
analyses of the structure of the neutrino flavor mixing. 
The analyses are based on the three-flavor mixing framework without 
sterile neutrinos and utilize the hints from solar 
and atmospheric neutrino observations as well as that from mixed dark 
matter cosmology. It will be demonstrated that the features of the 
analysis is sharply distinguished by the two cases (I) with and (II)
without dark matter mass scale, and by whether one (or two) 
mass is dominant (OMD) or the three states are almost degenerate (ADN). 
The global features of the neutrino mixing is illuminated for these 
different mass patterns.

\end{abstract}


In this talk, I summarize the progress in our understanding of the 
structure of lepton mixing matrix based on the analyses of neutrino 
mixing hinted by various experiments. I will put special emphasis 
on the fact that very weakly model-dependent approach exists toward 
determining patterns of neutrino mixing matrix. Probably, {\it now} 
is the time for us to start to think about the lepton flavor mixing, 
a fascinating subject which may guide us beyond the standard model 
of particle physics. 

Usually the standard electroweak theory is formulated by giving 
neutrinos zero mass. Do we have any experimental indication which 
is contradictory to the simplest possibility? I believe that the 
answer is yes. We do have three hints, whose two may be called as 
the ``direct experimental evidence'', while whose last is an 
intriguing suggestion from cosmology. The first of the former 
category is the energy-dependent modulation of the solar neutrino 
energy spectrum observed by four different experiments, the chlorine, 
the Kamiokande II-III, GALLEX and SAGE \cite {solar}. 
It is now very difficult to 
believe that the data from these four experiments can be reconciled 
with some sensible modification of the standard solar model. It is 
due to the low $^{71}$Ga rate which is well below 100 SNU and the 
missing (or strongly suppressed) $^7$Be neutrino situation enforced 
by the chlorine and the Kamiokande experiments. 

The second is the anomaly in the $\nu_{\mu}/\nu_e$ ratio in the 
atmospheric neutrinos observed by the Kamiokande, IMB and Soudan 2 
experiments \cite {atmospheric}. In spite of the fact that the 
anomaly is not observed by NUSEX and the Frejus detectors 
\cite {nusex}, the evidence in the Kamiokande 
experiment is so impressive that forces us to take it seriously. 
The existence (or not) of the anomaly will be checked by the 
well-operating Super-Kamiokande and the relevant region of the 
oscillation parameters will be probed by the planned long-baseline 
neutrino oscillation experiments. I note that the Japanese project, 
KEK $\rightarrow$ Super-Kamiokande, has already been funded and will 
run in early 1999. 

The third hint comes from the mixed (=hot and cold) dark matter 
cosmology \cite {chdm23}. It is one of the rival models which can 
account for the 
structure formation in the universe in a way consistent with the 
COBE observation. The neutrinos with masses of the order of $\sim$ 
10 eV provides the best candidate for hot component of the dark
matter. In fact, they are the only candidate particles that are 
known to exist in nature. We have to mention that the mixed dark 
matter cosmology itself cannot be regarded as compelling evidence 
for dark matter mass neutrinos. 

The important point which I am going to stress in this talk is that 
the existence, or nonexistence, of neutrinos with dark matter mass 
scale sharply distinguishes the features of the analysis of the 
neutrino mixing. I hope that clarification of this point finally 
allows us to obtain unambiguous criteria of how the (non-) existence 
of the dark matter mass neutrinos can be signaled experimentally. 
It is the ultimate goal of our investigation that the complete 
classification of the neutrino mixing pattern constrained by various 
experiments eventually leads us to the understanding of the physics 
of lepton flavor mixing. 

It is useful to classify neutrino mass patterns into I with and II 
without dark matter mass neutrinos. It is then important to classify 
the case I into two classes: 
\begin{flushleft} 
\hspace{0.5cm}I-OMD: One-Mass Dominance in type I 
(1 or 2 species of dark matter mass neutrinos)\\
\hspace{0.5cm}I-ADN: Almost Degenerate Neutrinos 
(all of which have dark matter mass)\\
\end{flushleft}
Then the case II may be denoted as II-ADN because there is no 
natural reason, besides dark matter cosmology, for expecting 
either intermediate or larger neutrino masses. So now the three types 
of mass pattern, I-OMD, I-ADN, and II-ADN, are on your shopping list. 

Let us define more precisely what we mean by I-OMD and 
I$\cdot$II-ADN to make our subsequent analyses unambiguous. 
By I-OMD we imply the mass pattern 
(a) $m_3^2 \gg m_1^2 \approx m_2^2$ 
where $\Delta m_{13}^2 \approx \Delta m_{23}^2 \sim 10$eV$^2$ and 
much larger than $\Delta m_{12}^2$, or the pattern 
(b) $m_3^2 \ll m_1^2 \approx m_2^2$ where 
$|\Delta m_{13}^2| \approx |\Delta m_{23}^2|$ $\sim$10eV$^2$. 
In the latter case (b) the term OMD is not quite correct; 
it is the two-mass dominant case. But we employ this terminology 
because it is becoming popular. Without matter effect the neutrino 
oscillation phenomenon itself does not distinguish between 
(a) and (b). The smaller $\Delta m_{12}^2$ can be taken as either 
$\Delta m_{12}^2 \sim 10^{-6}$eV$^2$, or 
$\Delta m_{12}^2 \sim 10^{-2}$eV$^2$, corresponding 
respectively to the solar neutrino and the atmospheric neutrino 
anomalies. Similarly we assume that the degeneracy in I$\cdot$II-ADN 
mass pattern is due to the solar and the atmospheric neutrino mass 
scales. Thus, the mass scales involved in the mass patterns which 
will be used in our following analysis entirely comes from one of 
the hints for neutrino masses that we mentioned before. 

We discuss in the following how these three neutrino mass patterns, 
I-OMD, I-ADN, and II-ADN, can be constrained by various experiments. 
We examine, one by one, (1) reactor and accelerator experiments, 
(2) solar neutrino experiments, 
(3) atmospheric neutrino experiments, and 
(4) double $\beta$ decay experiments. 
While detailed model building of the mixed dark matter 
cosmology may imply additional constraints, we rely on none of them 
quantitatively, except for the order of magnitude scale $\sim$10eV of 
dark matter mass. We believe that it is a reasonable attitude because 
systematic errors of the cosmological observations are difficult to 
estimate in most cases. 

We use the standard form of Cabibbo-Kobayashi-Maskawa quark mixing 
matrix

\begin{eqnarray}
U=\left[
\matrix {c_{12}c_{13} & s_{12}c_{13} &  s_{13}e^{-i\delta}\nonumber\\
-s_{12}c_{23}-c_{12}s_{23}s_{13}e^{i\delta} &
c_{12}c_{23}-s_{12}s_{23}s_{13}e^{i\delta} & s_{23}c_{13}\nonumber\\
s_{12}s_{23}-c_{12}c_{23}s_{13}e^{i\delta} &
-c_{12}s_{23}-s_{12}c_{23}s_{13}e^{i\delta} & c_{23}c_{13}\nonumber\\}
\right],
\label{eqn:CKM}
\end{eqnarray}
for the neutrino mixing matrix throughout the following analyses.
Its use became rather standard also in the analysis of the neutrino 
mixing. 

\vspace {0.5cm}
\noindent
{\large(1) Accelerator and reactor experiments}

The feature of the constraints from the accelerator and the reactor 
experiments markedly differ between mass patterns I-OMD and 
I$\cdot$II-ADN. By now it is well known that tremendous amount of 
constraints exist for the I-OMD case, but only a very mild one for 
I$\cdot$II-ADN. The clear recognition of these points began with 
the works in \cite{mina,BBGK}, and the bounds on the mixing parameters 
are worked out quantitatively in detail by Fogli, Lisi and Scioscia 
\cite{FLS} by using the explicit three-flavor angle definition 
as ours. The constraint for the I-OMD case is schematically drawn on 
$s_{23}^2-s_{13}^2$ plane in Fig. 1.

\vspace {0.5cm}
\noindent
{\large(2) Solar neutrino experiments}

The constraints from solar neutrino observation are again much 
milder for the mass pattern I$\cdot$II-ADN. It is only quite 
recently that the constraints for this case with the MSW mechanism 
has been worked out \cite{FLM} in a full three-flavor setting under 
the assumption of the $\Delta m^2$-hierarchy ($\neq$ mass hierarchy). 
They found, quite remarkably, that the small-$s_{12}$ and the 
large-$s_{12}$ solutions found in the two-flavor analyses are 
merely a part of the edge at $s_{13}=0$ in much broader 90\% CL 
allowed region on $s_{12}^2-s_{13}^2$ plane. Namely, the two 
small-$s_{13}$ solutions fuse into a single one at relatively 
large value of $s_{13}^2$ at around $s_{13}^2=0.33$, and the 
solution survives up to $s_{13}^2\simeq 0.6$. At the moment, 
the result obtained in \cite{FLM} constitutes our best knowledge 
of the three-flavor analysis of the solar neutrino constraint 
for the case I$\cdot$II-ADN. 

In the mass pattern I-OMD one can place stronger constraints by 
using the solar neutrino observation. One can show quite generally 
that a severe bound exists for $s_{13}^2$, $s_{13}^2 \lsim$ 
a few \%. It can be seen in Fig. 1, though schematically, that 
with dark matter mass neutrinos of OMD type mass pattern the 
accelerator and the reactor data constrain $s_{13}^2$ either 
less than a few \% level or very close to 1. Then, the useful 
formula \cite{formula}
\begin{equation}
\label{eq:formula}
P^{(3)}(\nu_e\rightarrow\nu_e) = c_{13}^4 P^{(2)}(\nu_e\rightarrow\nu_e)
+ s_{13}^4
\end{equation}
which relates the electron neutrino survival probabilities in the 
three- and the two-flavor frameworks implies that the large-$s_{13}$ 
branch is not consistent with the solar neutrino deficit. 
Notice that (\ref{eq:formula}) holds also for the vacuum neutrino 
oscillation and hence the conclusion prevails in the vacuum 
oscillation solution. Thus, the lower two domains in Fig. 1 are the 
allowed region by the terrestrial experiments and the solar neutrino 
observation for the case of I-OMD. 

\vspace {0.5cm}
\noindent
{\large(3) Atmospheric neutrino experiments}

The constraint from the atmospheric neutrino experiments gives rise 
to quite different allowed region on $s_{13}^2-s_{23}^2$ plane for 
the I-OMD and I$\cdot$II-ADN cases. In the case of I$\cdot$II-ADN 
mass pattern the allowed region is a $\Gamma$ (gamma)-shaped region 
on $s_{13}^2-s_{23}^2$ plane as drawn, again schematically, in 
Fig. 2. Despite the uncertainty we mention in the 
footnote\footnote[1]
{Unlike the case of the solar neutrino analysis the theoretical 
analysis of the atmospheric neutrino anomaly seems to be 
plagued with great uncertainties. It is realized by several 
researchers who tried to reproduce the features of the Monte-Carlo 
simulation (without neutrino oscillation) by Kamiokande group 
that it is quite difficult to achieve the goal. It might be due to 
the fact that theorists neither know the detector performance, 
nor have a chance of looking into the details of the Kamiokande
Monte-Carlo simulation (including its structure in energy-zenith 
angle plane) which cannot be read off from the published data. 
I must emphasize that unless one is able to reproduce the Kamiokande 
Monte-Carlo result with full use of the realistic neutrino flux, 
the reliability of the detailed shape of the obtained allowed region 
is subject to the great uncertainties. } 
this global feature seems to hold irrespective of the the details 
of the analysis. On the other hand, in the case of I-OMD the 
atmospheric neutrino anomaly must be taken care of by either 
almost pure 
$\nu_{\mu} \rightarrow \nu_{\tau}$ or $\nu_{\mu} \rightarrow \nu_e$ 
oscillations, if it is entirely due to the neutrino oscillation. 
It is because the constraints from accelerator and reactor experiments 
are so restrictive as to confine the allowed mixing parameters into 
the almost pure two-flavor channel. 
(We must bear in mind, however, that the full three-flavor analysis of 
the atmospheric neutrino data in this context has not been done so far.) 
By these simple considerations one can easily conclude that the relevant 
regions for accounting for the atmospheric neutrino anomaly in the I-OMD 
mass pattern is the upper horizontal region 
$(\nu_{\mu} \rightarrow \nu_{\tau})$ and the lower-left corner 
($\nu_{\mu} \rightarrow \nu_e$) on $s_{13}^2-s_{23}^2$ plane, 
as indicated in Fig. 1. 

I note that the allowed regions for the cases of I-OMD and 
I$\cdot$II-ADN are so different (they do not overlap!) is essentially 
due to the artifact of our definition of the mixing angle in the 
restricted framework of three-flavor mixing. If we work with an 
extended framework of I-OMD plus one sterile neutrino with the label 
0 (zero) the mixing angles corresponding to 
$\theta_{13}$ and $\theta_{23}$ in the I$\cdot$II-ADN scenario 
are ``$\theta_{02}$'' and ``$\theta_{12}$'', respectively. 

\vspace {0.5cm}
\noindent
{\large(4) Double $\beta$ decay experiments}

So far we have discussed the so called the ``$\Delta m^2$ physics'' 
in which only the mass-squared differences, not their absolute values, 
are the relevant quantities. This is the reason why our foregoing 
discussions did not distinguish between the mass patterns I-ADN and 
II-ADN, and between I-OMD (a) ($m_3^2 \gg m_1^2 \approx m_2^2$) 
and I-OMD (b) ($m_3^2 \ll m_1^2 \approx m_2^2$). 
Let us now address the experimental quantity that can 
distinguish these two cases. It is the neutrinoless double $\beta$ 
decay experiments by which one can constrain the Majorana neutrinos. 

Within the framework of I-ADN mass pattern  $<m_{\nu e}>$, the 
observable in neutrinoless double $\beta$ decay, can be greatly 
simplified. To understand the point let me remind you the 
generic expression of $<m_{\nu e}>$, 
\begin{equation}
<m_{\nu e}> = \left\vert c_{12}^2c_{13}^2 m_1 e^{-i(\beta+\gamma)}
+ s_{12}^2c_{13}^2 m_2 e^{i(\beta-\gamma)}
+ s_{13}^2 m_3 e^{2i(\gamma-\delta)},
\right\vert
\label{eq:beta1}
\end{equation}
where $\beta$ and $\gamma$ are the extra CP-violating phases
characteristic to Majorana neutrinos. In the I-ADN case one can 
ignore mass differences against the nearly degenerate neutrino 
masses themselves. In the CP-conserving case, as it is most 
obvious, the cancellation can take place depending upon the 
patterns of CP parities. Let us take the convention that 
$\eta_1= +$ and denote them collectively as 
$(\eta_1, \eta_2, \eta_3)\equiv
(1,e^{2i\beta},e^{i(\beta+\gamma+2\delta)}) =(+ + -)$ etc.
Then,
\begin{equation}
\label{eq:beta2}
r \equiv\frac{<m_{\nu e}>}{m}=
\left\{
\begin{array}{ll}
1 \;\;\;& \mbox{for} \;\;\;\;(+++) \\
\left\vert 1-2s_{13}^2 \right\vert\;\;\;& \mbox{for} \;\;\;\;(++-) \\
\left\vert 1-2s_{12}^2c_{13}^2 \right\vert\;\;\;& \mbox{for} \;\;\;\;(+-+) \\
\left\vert 1-2c_{12}^2c_{13}^2\right\vert \;\;\;& \mbox{for} \;\;\;\;(+--)
\end{array}
\right.
\end{equation}
Let us refer to the ratio $<m_{\nu e}>/m$ as $r$ hereafter.
  
An analysis based on the mixed dark matter scenario \cite {PS} 
gives degenerate dark matter neutrino mass in the range of 
2.3 eV $\lsim m_{\nu} \lsim$ 4.5 eV, which can be translated to 
0.15 $\lsim r \lsim$ 0.29 assuming the most stringent bound 
$<m_{\nu e}> \lsim $ 0.68 eV \cite {Moe}. 
Then, we have a constraint which can be drawn on 
$s_{12}^2 - s_{13}^2$ plane, as indicated in Fig. 3.  
The figure is for general CP-noninvariant cases and is based on our 
analysis done in Ref. \cite {MY}, to which we refer for details. 

In Fig. 3 the darker shaded region is the allowed region
with 90\% CL for the three-flavor MSW solution to the solar neutrino
problem obtained by Fogli et al. \cite {FLM} One can conclude that 
either the large-$s_{13}$ or the large-$s_{12}$ MSW solutions of the 
solar neutrino problem are preferred by the double $\beta$ decay 
constraint in the I-ADN scenario.

The double $\beta$ decay constraint matters also for the I-OMD mass 
pattern \cite {mina}. I briefly discuss the point. Generally speaking, 
the double $\beta$ decay constraint is difficult to achieve in the 
I-OMD (a) mass pattern because there is no chance of cancellation 
between two dark-matter masses. The constraint is, however, easily 
met in the small-$s_{13}$ region because the largest mass $m_3$
is multiplied by $s_{13}^2$. On the other hand, in I-OMD (b) mass 
pattern the cancellation is possible. In fact the cancellation is 
automatic in the large-$s_{13}$ region. Thus, the double $\beta$ 
decay constraints act in a rather nontrivial way in the I-OMD cases. 
For detail I refer Ref. \cite {mina}.  

In conclusion we have presented the results of our analysis of the 
neutrino mixing with and without mass hierarchy in the framework 
of three-flavor neutrinos without steriles. In the OMD case the 
mixing pattern is tightly constrained and is determined to be 
essentially unique when the double $\beta$ constraint is imposed. 
In the ADN mass pattern with dark matter scale we also obtain 
interesting constraints which prefer either the large-$s_{13}$ 
or the large-$s_{12}$ MSW solutions of the solar neutrino problem.

\clearpage
\vspace*{1cm}
\includegraphics{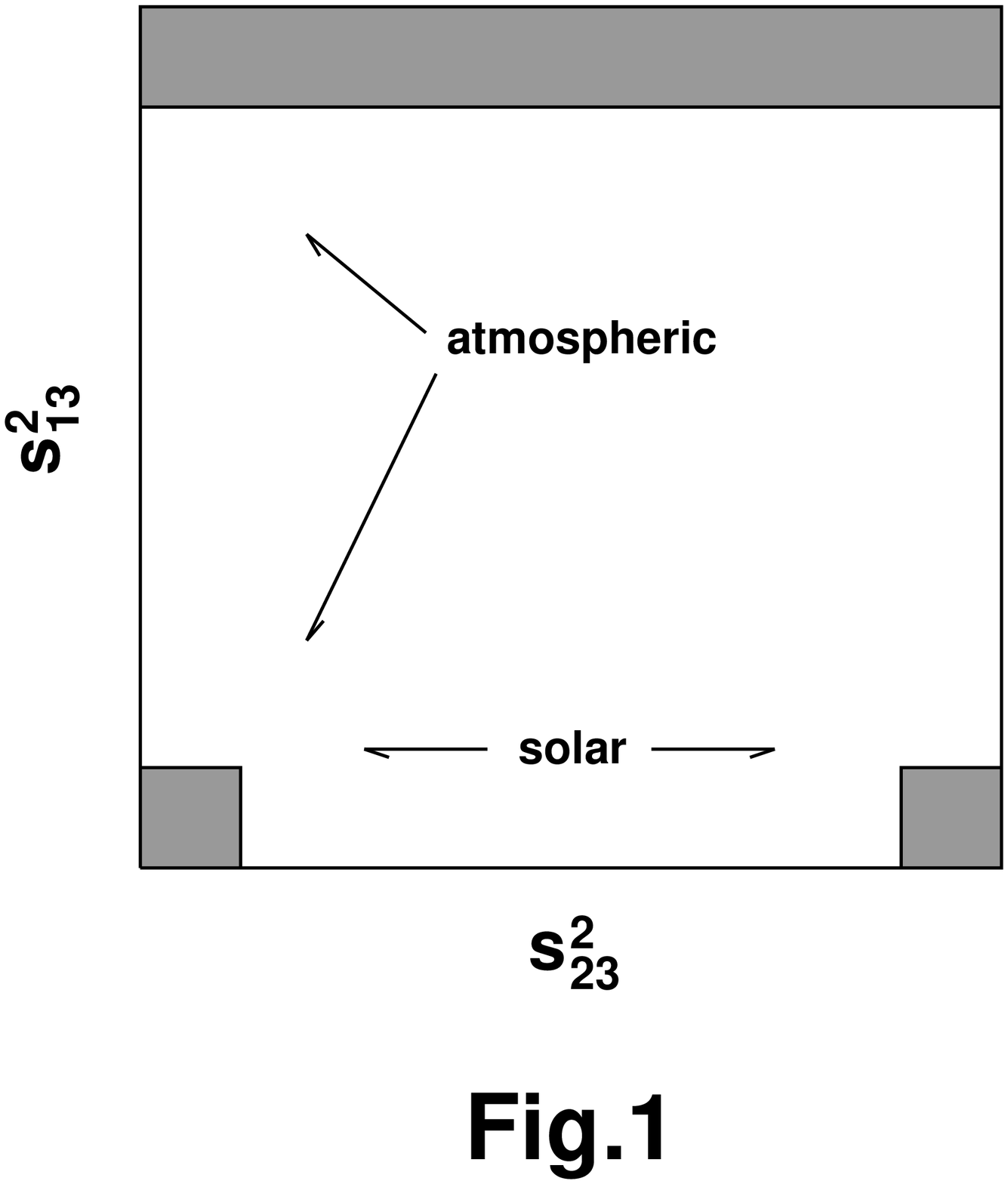}
\vspace{1cm}
\includegraphics{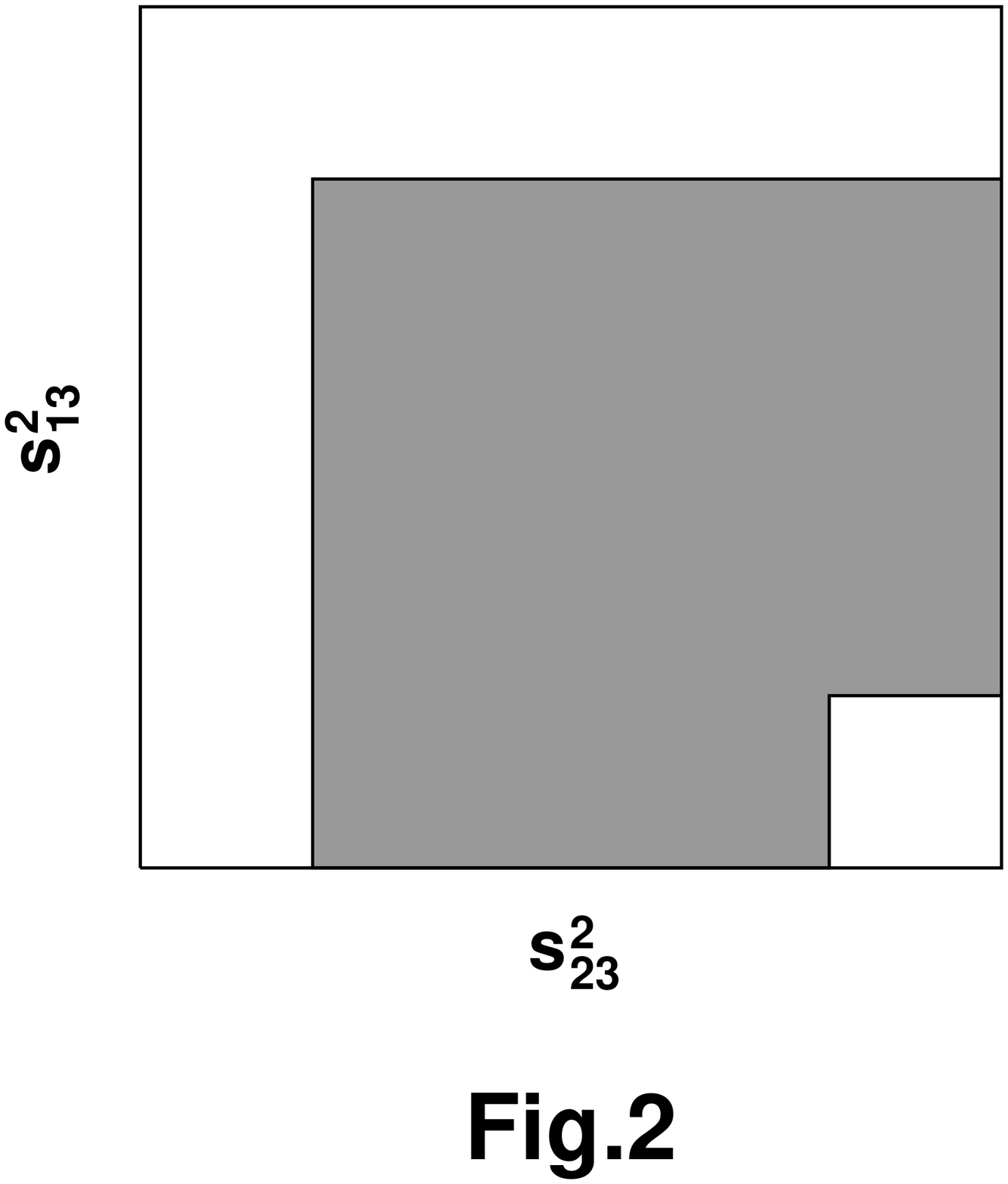}
\clearpage


%

%
%

\includegraphics{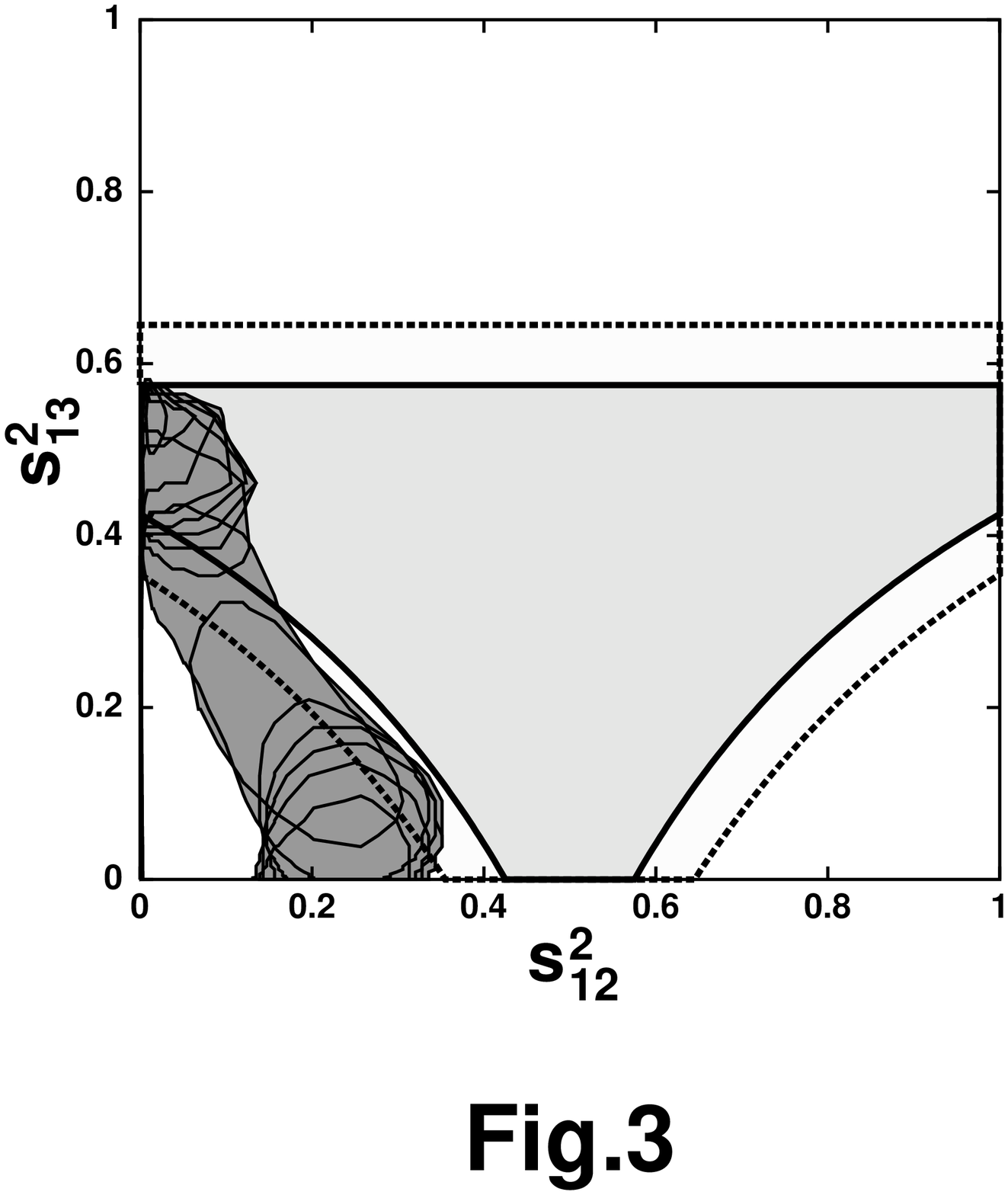}


\vspace*{12cm}
\hspace{2.2cm}
\begin{minipage}{11cm}
{\small
Fig.3: The areas inside the solid and the dashed lines are the 
allowed regions by the neutrinoless double $\beta$ decay experiments 
in general CP non-conserving cases where the CP violating phases 
$\beta$, $\gamma$ and $\delta$ are treated as unconstrained. 
The solid and the dashed lines are for $r<$ 0.15 which corresponds 
to $m$=4.5eV, and for $r<$ 0.29 which corresponds to $m$=2.3 eV, 
respectively. The darker shaded areas bounded by the thinner lines 
are the allowed regions with 90\% CL for the three-flavor MSW solution 
of the solar neutrino problem obtained by Fogli et al. \cite {FLM}.}
\end{minipage}


\begin{references}
\bibitem{solar}
B. T. Cleveland et al., Nucl. Phys. B (Proc. Suppl.) {\bf 38} (1995) 47;
Y. Suzuki, ibid {\bf 38} (1995) 54;
P. Anselmann et al., Phys. Lett. {\bf B285} (1992) 376; 
{\bf B314} (1993) 445; {\bf 327}, (1994) 377; {\bf B342}, (1995) 440;
J. N. Abdurashitov et al., Nucl. Phys. B (Proc. Suppl.) {\bf 38} (1995) 60.

\bibitem{atmospheric}
K. S. Hirata et al., Phys. Lett. {\bf B205} (1988) 416;
{\bf B280} (1992) 146;
Y. Fukuda et al., ibid {\bf B335} (1994) 237;
R. Becker-Szendy et al., Phys. Rev. {\bf D46} (1992) 3720;
M. C. Goodman, Nucl. Phys. B (Proc. Suppl.) {\bf 38} (1995) 337.

\bibitem {nusex}
M. Aglietta et al., Europhys. Lett. {\bf 8(7)} (1989) 611;
K. Daum et al., Z. Phys.  {\bf C66} (1995) 417.

\bibitem {chdm23}
J. A. Holtzman, Astrophys. J. Suppl. {\bf 71} (1989) 1;
J. A. Holtzman and J. R. Primack, Astrophys. J. {\bf 405} (1993) 428;
J. R. Primack, J. Holtzman, A. Klypin, and D. O. Caldwell,
Phys. Rev. Lett. {\bf 74} (1995) 2160;
K. S. Babu, R. K. Schaefer, and Q. Shafi,  Phys. Rev. {\bf D53} (1996)
606; D. Pogosyan and A. Starobinsky, astro-ph/9502019.

\bibitem {mina}
H. Minakata, Phys. Rev. {\bf D52} (1995) 6630; Phys. Lett. {\bf B356}
(1995) 61.

\bibitem {BBGK}
S. M. Bilenky, A. Bottino, C. Giunti, and C. W. Kim, Phys. Lett.
{\bf B356} (1995) 273.

\bibitem {FLS}
G. L. Fogli, E. Lisi, and G. Scioscia, Phys. Rev. {\bf D52} (1995) 5334.

\bibitem {FLM}
G. L. Fogli, E. Lisi, and D. Montanino,  Phys. Rev. {\bf D54} (1996) 2048.

\bibitem {formula}
C.-S. Lim, Proceedings of BNL Neutrino Workshop, February 1987, BNL 52079; 
X. Shi and D. N. Schramm, Phys. Lett. {\bf B283} (1992) 305; 
A. Yu. Smirnov, in {\it Frontiers of Neutrino Astrophysics}, Universal 
Academy Press, 1993.
 
\bibitem {PS}
D. Pogosyan and A. Starobinsky, in Ref. \cite {chdm23}.

\bibitem{Moe}
M. K. Moe, Nucl. Phys. B (Proc. Suppl.) {\bf 38} (1995) 36.

\bibitem {MY}
H. Minakata and O. Yasuda, TMUP-HEL-9610, hep-ph/9609276. 

\end{references}
\end{document}